\documentclass[11pt,twoside,onecolumn]{article}
\newcommand{\be}{\begin{eqnarray}}
\newcommand{\ee}{\end{eqnarray}}
\newcommand{\bra}[1]{\mbox{$\langle\, #1 \mid$}}

\newcommand{\ket}[1]{\mbox{$\mid #1\,\rangle$}}

\newcommand{\pro}[2]{\mbox{$\langle\, #1 \mid #2\,\rangle$}}
\newcommand{\expec}[1]{\mbox{$\langle\, #1\,\rangle$}}
\newcommand{\expecl}[1]{\mbox{$\left\langle\,
            \strut\displaystyle{#1}\,\right\rangle$}}

\title{On gravitational fluctuations and the semi-classical limit
in minisuperspace models}
\author{Roberto Casadio\thanks{e--mail: casadio@bo.infn.it},
 \\
{\em Dipartimento di Fisica, Universit\`a di
Bologna} \\
{\em and} \\
{\em Istituto Nazionale di Fisica Nucleare,
Sezione di Bologna, Italy}}
\begin{document}
%
%
\maketitle
\begin{abstract}
An attempt is made to go beyond the semi-classical approximation
for gravity in the Born-Oppenheimer decomposition of the wave-function
in minisuperspace.
New terms are included which correspond to quantum gravitational
fluctuations on the background metric.
They induce a back-reaction on the semi-classical background and can
lead to the avoidance of the singularities the classical theory
predicts in cosmology and in the gravitational collapse of compact
objects.
\end{abstract}
\raggedbottom
\setcounter{page}{1}
\section{Introduction}
The canonical quantization of highly symmetrical general relativistic
systems carried out in suitably chosen variables leads to the dynamics
being determined by the (super)Hamiltonian constraint
\cite{dewitt,wheeler} of the Arnowitt, Deser and Misner (ADM)
construction \cite{adm} in the space of functions of time called
{\em minisuperspace} \cite{dewitt,misner}.
Such an approach is particularly useful to investigate self-gravitating
quantized matter with gravity in the semi-classical regime.
One performs a Born-Oppenheimer (BO) decomposition of the wave-function
satisfying the Wheeler-DeWitt (WDW) equation into two parts \cite{bv}.
The first one represents a collective degree of freedom associated
with gravity ({\em slow} component) and, in the semi-classical
approximation, leads to an Hamilton-Jacobi (HJ) equation for the
gravitational degree of freedom;
the second part describes microscopic matter ({\em fast} component)
and satisfies a Schr\"odinger equation in the time defined by
semi-classical gravity.
\par
Alternative approaches have been attempted, such as the one in
Ref.~\cite{halliwell}, where, however, relative phases of matter and
gravity were incorrectly identified \cite{bv}, or which involve an
expansion in the Planck mass (see, {\em e.g.}, Ref.~\cite{kiefer,hu}).
The latter expansion is potentially dangerous, since it has been shown
that it can lead to violation of unitarity within the framework of
canonical quantization \cite{bfv} and to incorrect identification of
the background as an empty solution of Einstein equations
\cite{bfv,parentani}.
In the BO approach \cite{bv} the collective degree of freedom evolves
slowly because it is associated with the total mass of the system
which is (many) times the mass of each constituting matter quantum
(regardless of the latter being smaller than the Planck mass).
For these reasons we shall appeal to the approach introduced in
Ref.~\cite{bv} as the best suited for the purpose of analyzing the
semi-classical limit and shall not attempt at expanding the
total wave-function in the Planck mass.
\par
The BO approach was subsequently applied to two physical models of
general interest:
the gravitational collapse of a sphere of homogeneous dust in empty
space \cite{cv} and spatially homogeneous Universes \cite{bfv,bfv2}
(see also Ref.~\cite{shell} for collapsing shells).
For the former system the novel effect of non-adiabatic production
of matter has been studied with the analytical method of the
({\em adiabatic}) {\em invariants} for time dependent Hamiltonians
\cite{lewis,bfv} in Ref.~\cite{cfv}.
The same technique, supplemented by numerical simulations, has shown
the possibility of having an inflationary phase in the primordial
Universe which is driven by purely quantum fluctuations of the
inflaton and has finite duration \cite{fvv}.
In both cases there are one degree of freedom for gravity,
$R$ (related to the external radius of the sphere or the scale
factor of the Universe), and one degree of freedom for matter, $\phi$
(homogeneous scalar field).
The phase space is then the usual Friedmann-Robertson-Walker (FRW)
minisuperspace of the space-time metric
\be
ds^2=R\,\left[-d\eta^2+{d\rho^2\over 1-\epsilon\,\rho^2}+\rho^2\,
\left(d\theta^2+\sin^2\theta\,d\varphi^2\right)\right]
\ee
($\epsilon=0,\pm 1$ respectively for flat, spherical and hyperbolic
space and $\rho\ge 0$ for $\epsilon=0,-1$; $0\le \rho\le 1$ for
$\epsilon=1$) and homogeneous scalar matter.
The coordinates $(\eta,\rho,\theta,\varphi)$ define a comoving
reference frame and for the sphere of dust $\rho\le \rho_s$, where
$\rho_s$ is the (constant) comoving radius of the sphere.
\par
One must be careful in modeling dust with a scalar field, since the
latter indeed describes a perfect fluid with pressure equal to the
Lagrangian density,
$p\sim\frac{1}{2}\,(\dot\phi^2-\ell_\phi^{-2}\,\phi^2)$
\cite{madsen}.
If the scalar field has mass $m_\phi=\hbar/\ell_\phi$, then
$p$ oscillates with frequency $\sim 2/\ell_\phi$, {\em e.g.},
for $m_\phi\sim 10^{-27}\,$kg (the proton mass) this means a period
$T\sim 10^{-23}\,$s.
It is thus reasonable to approximate the actual pressure with its time
average over one period (that is, set $p=0$) provided the
radius $R$ does not change appreciably on the time scale $T$
({\em quantum adiabatic approximation} for the state of the scalar
field).
Moreover, this adiabatic approximation becomes exact in the classical
limit for $\phi$, as can be seen by taking $\hbar\to 0$ with $m_\phi$
held fixed ($\ell_\phi\to 0$ and $T\sim \ell_\phi$ vanishes),
and a mode of the homogeneous massive scalar field can be identified
with dust.
In fact in Ref.~\cite{cv} it was verified that in this approximation
one recovers the classical Oppenheimer-Snyder (OS) model \cite{OS}.
\par
A major restriction in Refs.~\cite{bv,bfv,cv,bfv2,shell,cfv,fvv} is
that quantum fluctuations of the gravitational degree of freedom
were suppressed {\em a priori} and $R$ was approximated by a
classical trajectory $R_c(\eta_c)$ ($\eta_c$ being the
{\em conformal time} associated with that trajectory) which, in turn,
was determined solely by the matter content.
The aim of the present notes is to allow the variable $R$ to have
quantum fluctuations around the classical trajectory and modify the
expressions of the general formalism \cite{bfv} accordingly.
Of course, it would be much more interesting to allow for inhomogeneous
fluctuations, but this would inevitably render the system intractable
analitically and is left for future developments.
As a by-product, we will see that one can have a significant
back-reaction of the gravitational quantum dynamics on the
semi-classical trajectory.
This affects the singularity classical General Relativity generically
predicts in cosmology and as the final state of a collapsing
body (see \cite{hawking} and Refs. therein).
\par
The possibility of avoiding space-time singularities in a quantum
theory has been studied for a long time and the literature on this
topic is wide.
Here we only refer to two approaches:
\begin{enumerate}
\item
in quantum field theory in curved space-time (see, {\em e.g.},
Ref.~\cite{birrell}) gravity is described by a classical background
on which quantum matter fields propagate.
In Ref.~\cite{parker} it was found that there are states of matter for
which the Universe admits a minimum non-zero scale factor, provided
the number of particles is not conserved;
\item
it was suggested that canonical quantization of the gravitational
degrees of freedom could bypass the cosmological singularity
\cite{wheeler2}.
In Ref.~\cite{misner2} the constraints were implemented before
quantizing and one ended up with quantized gravitational degrees of
freedom only.
In this case no significant change in the classical behaviour was
found.
\end{enumerate}
It is a trivial observation that in a quantum theory a point-like
singularity is meaningless since it would violate Heisenberg's
principle.
What we shall show in the proposed approach is that one expects
the singularity is avoided under a broad assignment of initial
conditions.
In fact the semi-classical approximation breaks down before
the point-like singularity is reached (but within the adiabatic
approximation for the gravitational degree of freedom) and the very
concept of a trajectory loses its meaning at a value of $R$ which
can be appreciably big (in a sense that will be specified later).
\par
The plan of the paper is as follows.
In the next Section quantum gravitational fluctuations are treated
in the standard BO formalism for the FRW minisuperspace and it is
shown that their energy cannot always be neglected with respect to
the energy of matter.
In Section~\ref{EHJ} the energy of such fluctuations is incorporated
in a modified semi-classical HJ equation which is solved under certain
approximations.
Such approximations are then analyzed to determine the range of
validity of the solutions.
In Section~\ref{applications} some conclusions are drawn
for cosmological models and for the collapse of homogeneous spheres of
dust.
Finally in Section~\ref{conc} the results are summarized and commented.
we shall use units in which $c=1$, $\kappa=8\,\pi\,G_N$,
$\ell_p=\sqrt{\hbar\,\kappa}$ is the Planck length.
\setcounter{equation}{0}
\section{Quantum gravitational fluctuations in the BO approach}
\label{rev}
Let us start directly from the WDW equation in the minisuperspace of
the two variables $R$ and $\phi$ (for a derivation from first principles
see \cite{dewitt}) with a convenient {\em operator ordering} in the
gravitational kinetic term \cite{cv}:
\be
\left[\hat H_G+\hat H_M\right]\,\Psi\equiv
{1\over 2}\,\left[{\kappa\hbar^2}\,
\frac{\partial^2}{\partial R^2} {1\over R}
-{\epsilon\over\kappa}\,R
-{\hbar^2\over R^3}\,
{\partial^2\over\partial\phi^2}
+{1\over\ell_\phi^2}\,\phi^2\,R^3
\right]\,\Psi(R,\phi)
=0
\ .
\label{wdw}
\ee
The wave-function $\Psi$ can be expressed in the factorized form
$\Psi(R,\phi)=R\,\psi(R)\,\chi(\phi,R)$ which, after multiplying on
the LHS of Eq.~(\ref{wdw}) by $\chi^\ast$ and integrating over the
matter degrees of freedom, leads to the equation for the gravitational
part \cite{bfv}
\begin{eqnarray}
& &{1\over2}\left[
\left({\kappa\hbar^2}\,{\partial^2\over\partial R^2}
-{\epsilon\over\kappa}\,R^2\right)
+{1\over\pro{\tilde\chi}{\tilde\chi}}\,
\bra{\tilde\chi}\,
\left({{\hat\pi_{\phi}}^2\over R^2}
+{1\over \ell_\phi^2}\,\phi^2\,R^4\right)
\,\ket{\tilde\chi}\right]\,\tilde\psi
\nonumber\\
& &\equiv\left[\hat H_{_G}\,R+R\,\expec{\hat H_{_M}}
\right]\,\tilde\psi
=\frac{\kappa\hbar^2}{2 \pro{\chi}{\chi}}
\langle \chi |
\frac{\stackrel{\leftarrow}{\partial}}{\partial R}
\left( 1 - \frac{\ket{\chi} \bra{\chi}}{\expec{\chi|\chi}} \right)
\frac{\partial}{\partial R} | \chi \rangle
\,\tilde\psi
\ .
\label{wdw_g}
\end{eqnarray}
The scalar product $\pro{\chi}{\chi}\equiv
\int d\phi\,{\chi}^\ast(\phi,R)\,\chi(\phi,R)$
and
\be
\begin{array}{lcr}
\psi=e^{-i\,\int^R A(R')\,dR'}\,\tilde\psi &
\ \ \ \ \ \ &
\chi=e^{+i\,\int^R A(R')\,dR'}\,\tilde\chi
\ ,
\end{array}
\ee
with $A\equiv -i\,\pro{\chi}{\chi}^{-1}\,
\bra{\chi}\,{\partial_R}\,\ket{\chi}
\equiv -i \,\expec{\partial_R}$.
If we now multiply Eq.~(\ref{wdw_g}) by $\tilde\chi$ and subtract it
from Eq.~(\ref{wdw}) we obtain the equation for the matter
function $\tilde\chi$ \cite{bfv}
\begin{eqnarray}
\tilde\psi\,R
\,\left[\hat H_{_M}-\expec{\hat H_{_M}}\right]\,\tilde\chi
+{\kappa\hbar^2}\,
\left({\partial\tilde\psi\over\partial R}\right)\,
{\partial\tilde\chi\over\partial R}
={\kappa\hbar^2\over2}\,\tilde\psi\,\left[
\expecl{{\partial^2\over\partial R^2}}-
\frac{\partial^2}{\partial R^2}\right]\,\tilde\chi
\ .
\label{wdw_m}
\end{eqnarray}
The Eqs.~(\ref{wdw_g}) and (\ref{wdw_m}), as well as the WDW equation,
are exact, in the sense that no approximation has been assumed yet for
the wave-functions $\tilde\chi$ and $\tilde\psi$, and contain no time
variable.
\par
A way one can introduce the time is by taking the semi-classical
limit for gravity \cite{banks,brout,bv}.
In order to do so, first one needs to neglect the RHS's of
Eqs.~(\ref{wdw_g}) and (\ref{wdw_m}) which are related to quantum
transitions among different semi-classical trajectories \cite{bfv}.
As usual \cite{bv,bfv}, we shall check the consistency of all
approximations once the solutions to the semi-classical equations
have been obtained (see Section~\ref{conditions}).
In fact, it is not necessary (nor possible, in general) to prove
that the RHS's are small from the onset, but it is sufficient to show
{\em a posteriori\/} that they are negligible for the cases considered.
Then one writes a semi-classical (WKB) approximation for the wave
function $\tilde\psi$
\be
\psi_c =  {1\over\sqrt{-P_c}}\,
e^{+{i\over\hbar}\,\strut\displaystyle\int P_c\,dR_c}
\ ,
\label{wkb}
\ee
where
\be
P_c=-{1\over \kappa}\,{\partial R_c\over\partial\eta_c}=
-{1\over\kappa}\,
\sqrt{2\,\kappa\,R_c\,\expec{\hat H_M}-\epsilon\,R_c^2}
\
\label{pi}
\ee
is the canonical momentum conjugated to $R$ in the classically allowed
region $\expec{\hat H_M}>\epsilon\,R^2/2\,\kappa$
and the integral in the exponent is computed along the
(so far unspecified) semi-classical trajectory $R=R_c(\eta_c)$
with momentum $P=P_c(\eta_c)$.
Moreover, the derivatives with respect to the conformal time $\eta_c$
are defined according to Eq.~(\ref{pi}) as
\be
{\partial\over\partial \eta_c}
\equiv -\kappa\,\psi_c\,P\,{\partial\over\partial R}
=-\kappa\,P_c\,\left.{\partial\over\partial R}\right|_{R_c}
\ ,
\label{d_eta}
\ee
where the last step follows from $\psi_c$ having support only for
$R\sim R_c$, $\eta\sim\eta_c$.
\par
Upon substituting $\tilde\psi=\psi_c$ into Eq.~(\ref{wdw_g}),
the gravitational equation finally reduces to the
semi-classical HJ equation
\be
\psi_c\,\left[-{1\over2\kappa}\,\left({d R\over d\eta}\right)^2
-{\epsilon\over2\kappa}\,R^2+R\,\expec{\hat H_M}\right]
=
-{1\over2\kappa}\,\left({d R_c\over d\eta_c}\right)^2
-{\epsilon\over2\kappa}\,R_c^2+R_c\,\expec{\hat H_M}=0
\ ,
\label{hj}
\ee
which can now be used to determine $R_c$ explicitly once
$\expec{\hat H_M}$ is given.
\par
It is important to note that the semi-classical regime is not defined
simply as the limit $\hbar\to 0$, but rather by a specific choice of
the wave-function $\psi_c$.
For instance, with $P_c$ given by Eq.~(\ref{pi}),
$\expec{\hat H_M}=N_\phi\,m_\phi$ and a constant
radial number density of scalar quanta $N_\phi>0$
(in practice this is the statement of the quantum adiabatic
approximation), one has (see, {\em e.g.}, Ref.~\cite{stephani})
\be
R_c=N_\phi\,{\ell_p^2\over \ell_\phi}\times
\left\{\begin{array}{ll}
(\cosh\eta_c-1) & \epsilon=-1 \\
\\
\eta_c^2/2 & \epsilon=0 \\
\\
(1-\cos\eta_c) & \epsilon=+1
\ ,
\end{array}\right.
\label{kc}
\ee
that is the usual FRW cosmological models for increasing $\eta_c$
or the OS model of gravitational collapse for decreasing $\eta_c$
(the classical singularity occurs at $\eta_c=0$ in both cases).
\par
Substituting $\tilde\psi=\psi_c$ in Eq.~(\ref{wdw_m}) gives
the Schr\"o\-din\-ger equation
\be
i\,\hbar\,{\partial\chi_s\over\partial\eta_c}
={1\over 2}\,
\left[-{\hbar^2\over R_c^2}\,{\partial^2\over\partial\phi^2}
+{1\over \ell_\phi^2}\,R_c^4\,\phi^2\right]\,\chi_s
\
\label{schro}
\ee
for the rescaled matter function
$\tilde\chi=\chi_s\,\exp\{(i/\hbar)\,\int^\eta_c R_c\,d\eta_c'\,
\expec{\hat H_M}\}$.
We note in passing that the difference between $\chi_s$ and the
original $\chi$ amounts exactly to the phase factor between
eigenvalues of an hermitian invariant for the Hamiltonian in
Eq.~(\ref{schro}) \cite{lewis} and exact solutions of
Eq.~(\ref{schro}) \cite{bfv}.
The above Schr\"odinger equation together with the
HJ equation (\ref{hj}) was the starting point
for the results found in Refs.~\cite{cv,bfv2,shell,cfv,fvv}
and led to the conclusions briefly mentioned in the introduction.
\par
So far matter and gravity are determined by two equations of clearly
different types.
Suppose instead one defines
\be
\tilde\psi=\psi_c\,f
\ ,
\label{ewkb}
\ee
where $f=f(R)$ will encode quantum fluctuations around the trajectory
$R_c$ swept by $\psi_c$.
Then Eq.~(\ref{wdw_g}), again neglecting the RHS's (see
Section~\ref{conditions} for a detailed description of the difference
with the previous case), becomes an equation for $f$
($\ '\equiv\partial/\partial R$)
\be
&&\psi_c\,\left[{\kappa\hbar^2\over 2}\,\left(
{3\over 4}\,{{P'}^2\over P^2}-{P''\over2\,P}
-{P'\over P}\,{\partial\over\partial R}
+{\partial^2\over\partial R^2}\right)
+i\,\kappa\hbar\,P\,{\partial\over\partial R}\right]f
\nonumber \\
&&=\psi_c\,\left[
{\kappa\over 2}\,P^2+{\epsilon\over2\kappa}\,R^2
-R\,\expec{\hat H_M}\right]\,f
\ .
\label{ww}
\ee
Upon using the definition (\ref{d_eta}), the above takes the form of a
time-dependent Schr\"odinger equation
($\ \dot{ }\equiv\partial/\partial\eta$),
\be
i\,\hbar\,{\partial f\over\partial\eta_c}&=&
\psi_c\,{\kappa\hbar^2\over 2}\,\left[
{5\over 4\kappa^2}\,{{\dot P}^2\over P^4}
-{1\over2\kappa^2}\,{\ddot P\over P^3}
+{1\over\kappa}\,{\dot P\over P^2}\,{\partial\over\partial R}
+{\partial^2\over\partial R^2}\right]\,f
\nonumber \\
&&+\psi_c\,\left[-{\kappa\over 2}\,P^2
-{\epsilon\over2\kappa}\,R^2
+R\,\expec{\hat H_M}\right]\,f
\ .
\label{sc}
\ee
In the RHS, due to the factor $\psi_c$, the quantity $P$ is still
evaluated at the classical momentum and $\eta=\eta_c$.
Therefore one can simplify Eq.~(\ref{sc}) by making use of the
HJ equation (\ref{hj}) and obtains
\be
i\,\hbar\,{\partial f\over\partial\eta_c}=
{\kappa\hbar^2\over 2}\,\left[\left(
{5\over 4\kappa^2}\,{\dot P_c^2\over P_c^4}
-{1\over2\kappa^2}\,{\ddot P_c\over P_c^3}\right)\,f(R_c)
+{1\over\kappa}\,{\dot P_c\over P_c^2}\,
\left.{\partial f\over\partial R}\right|_{R_c}
+\left.{\partial^2 f\over\partial R^2}\right|_{R_c}\right]
\ ,
\label{sc1}
\ee
which is now the analogue of Eq.~(\ref{schro}) for gravity.
\par
Since Eq.~(\ref{sc1}) is involved, let us take the {\em classical
adiabatic limit} for $R_c$, to wit $|\dot R_c|\ll R_c\Rightarrow
|\partial^{n+1}R_c/\partial\eta_c^{n+1}|\ll |\dot R_c|$
for any integer $n>0$.
This approximation is not to be confused with the previously mentioned
quantum adiabatic limit on the state of $\phi$ and will be further
discussed in Section~\ref{conditions}.
From the definitions (\ref{pi}) and (\ref{d_eta}) it then follows that
one can neglect terms containing $\dot P_c$ and $\ddot P_c$.
Hence, only the last term survives in the RHS above and one finds
\be
i\,\hbar\,{\partial f\over\partial\eta_c}=
{\kappa\hbar^2\over 2}\,
\left.{\partial^2 f\over \partial R^2}\right|_{R_c}
\ ,
\label{sc0}
\ee
which resembles the non-relativistic equation for a free particle
of ``mass'' $1/\kappa$ and negative kinetic energy.
\subsubsection*{Plane waves}
Eq.~(\ref{sc0}) admits solutions in the form of plane waves,
\be
f_\lambda=
\exp\left\{
i\,{\ell_p^2\,\eta_c\over2\,\lambda^2}+i\,{R_c\over\lambda}\right\}
\ ,
\label{f_w}
\ee
where the $\lambda$'s are real numbers ($\lambda>0$ for left movers
and $\lambda<0$ for right movers in $R$ space).
One observes that, due to the ``wrong'' sign mentioned above,
the energy (conjugated to the proper time $d\tau=R_c\,d\eta_c$)
associated with each mode of wavelength $\lambda$,
\be
E_\lambda=-{\hbar\over 2\,R_c}\,{\ell_p^2\over\lambda^2}
\ ,
\label{Ela}
\ee
is negative.
Although disturbing at first sight, this is in agreement with gravity
contributing negative amounts to the total (super)Hamiltonian
\cite{opposite}.
\subsubsection*{Exponential waves}
For imaginary $\lambda=i\,l$, from Eq.~(\ref{f_w}) one obtains a new
set of solutions given by
\be
f_l=\exp\left\{-i\,{\ell_p^2\,\eta_c\over2\,l^2}+{R_c\over l}\right\}
\ ,
\ee
whose energies are positive,
\be
E_l={\hbar\over 2\,R_c}\,{\ell_p^2\over l^2}
\ .
\label{El}
\ee
Of course the amplitude of the above solutions increase with $R_c$
for $l>0$ and decrease with $R_c$ for $l<0$, thus signalling
an instability.
In fact, these modes can be related to the tunneling of $R$ across
classically forbidden regions, since $i\,P\,(\ln f_l)'$ is real if
$P$ is imaginary.
For this reason, one cannot superpose solutions from the two sets
$\{f_\lambda\}$ and $\{f_l\}$ if the classical limit for $R$ has
to make sense \cite{forth} (due to the factor $\psi_c$ either $R$
is in a classically allowed region or it is not).
\par
In the following we will only consider real values for $\lambda$.
\subsubsection*{Full solutions}
The full gravitational state corresponding to the modes $f_\lambda$
found above are given by
\be
\tilde\psi_\lambda\equiv\psi_c\,f_\lambda= \psi_c\,
\exp\left\{
i\,{\ell_p^2\,\eta_c\over2\,\lambda^2}+i\,{R_c\over\lambda}\right\}
\ ,
\label{full}
\ee
where the weight $\psi_c$ ensures that
$\int dR\,\tilde\psi_\lambda^\ast\,\hat O\,\tilde\psi_\lambda=
(\hat O\,f_\lambda)(R_c)$ for every operator
$\hat O(R,\partial/\partial R)$.
Then the general solution to Eq.~(\ref{sc0}) is a superposition of
the form $\sum c_\lambda\,f_\lambda(R_c)$ and the total energy
associated to quantum gravitational fluctuations is given by
\be
E_f=-{\hbar\,\ell_p^2\over 2\,R_c}\,
\sum\limits_\lambda\,{|c_\lambda|^2\over\lambda^2}
\ ,
\label{Ef}
\ee
where the $c_\lambda$ are normalization coefficients.
\par
It is clear from Eq.~(\ref{Ef}) that $|E_f|$ can be very large,
depending on the modes $f_\lambda$ which are included.
In particular, $|E_\lambda|\ll\expec{\hat H_M}$ only for
\be
\lambda^2\gg\lambda_c^2\equiv
{\hbar\over 2}\,{\ell_p^2\over R_c\,\expec{\hat H_M}}
\ .
\label{UV}
\ee
An interesting observation is that $\lambda_c$ is time-dependent
(via $R_c$) and, in the quantum adiabatic approximation for $\chi_s$,
the term $\expec{\hat H_M}=N_\phi\,m_\phi$ is constant \cite{cv}
and one has
\be
\lambda_c^2={\ell_p^2\,\ell_\phi\over 2\,R_c\,N_\phi}
\ .
\ee
For an expanding universe in which $R_c$ increases in time without
bounds, $\lambda_c$ will eventually vanish after it had been as big
as possible in the far past.
On the other hand, for the case of a collapsing sphere of dust with
monotonically decreasing $R_c$, $\lambda_c$ will diverge and, no matter
how long are the wavelengths of the initial gravitational
fluctuations, $|E_f|$ will overcome $\expec{\hat H_M}$ before the
sphere reaches the classical singularity $R_c=0$.
\par
We thus arrive at the following paradoxical conclusion.
Our equations show that there are quantum gravitational fluctuations
which can be generally associated to the classical solutions $R_c$ in
Eq.~(\ref{kc}).
The energy of such fluctuations becomes inevitably larger then the
matter energy at certain times but, since it does not appear in the
HJ equation, the presence of gravitational fluctuations does not
affect the semi-classical motion in any way.
In the next Section we shall show that this paradox is due to an
incorrect identification of the semi-classical limit and propose
an approach to include the back-reaction of gravitational
fluctuations.
\setcounter{equation}{0}
\section{Improved BO approach}
\label{EHJ}
The aim of this Section is to propose a redefinition of the
semi-classical limit for gravity in minisuperspace which
includes the (negative) energy of the gravitational fluctuations
found in the previous Section into the HJ equation.
This amounts to treat the gravitational fluctuations as an extra
``matter'' contribution, in much the same fashion as is usually done
in perturbation theory around a fixed background in order to
compute the back-reaction on the metric
(see, {\em e.g.}, \cite{birrell}).
The main advantage of the BO decomposition \cite{bv,bfv} with respect
to the latter approach is that we now derive such a description from a
(supposedly more fundamental) unitary quantum theory (the WDW equation
(\ref{wdw})) together with explicit conditions for the semi-classical
approximation which should otherwise be deduced from external
principles.
In fact, this will give us (semi)classical trajectories $R_f$
corresponding to the matter content $\expec{\hat H_M}$ and the
gravitational state $f$ together with the consistency conditions
discussed in Section~\ref{conditions}.
\par
In order to simplify the analysis from now on we shall consider
one gravitational mode at a time and set
$\tilde\psi=\tilde\psi_\lambda$
so that the energy of the gravitational fluctuations is given by
$E_\lambda$ in Eq.~(\ref{Ela}).
With the above restriction, Eq.~(\ref{ww}) in the classical adiabatic
approximation $\dot R\ll R$ becomes
\be
-i\,\kappa\hbar\,\psi_c\,P\,{\partial f_\lambda\over \partial R}=
\psi_c\,\left[-{\kappa\over 2}\,P^2-{\epsilon\over2\kappa}\,R^2
+R\,\expec{\hat H_M}
-{\kappa\hbar^2\over 2\,\lambda^2}\right]\,f_\lambda
\ .
\label{scc}
\ee
The term of order $(\hbar/\lambda)^2$ survives in the semi-classical
limit only provided one allows for very short wavelengths, such that
$\ell_p/|\lambda|$ does not vanish for $\hbar\to 0$.
This is just the analogue of what is required for the expectation
value of the matter Hamiltonian ($\sim \hbar/\ell_\phi$),
to wit $\ell_\phi\sim\hbar$.
The condition $|\lambda|\sim\ell_p$, in turn, would refer to
a fully quantum theory of gravity, if $\lambda$ is interpreted as a
spatial wavelength, and one might prefer to place a ultra-violet
cut-off $\Lambda\ge \ell_p$ for the values of $\lambda$.
We prefer to stick to a more euristic attitude and assign a physical
meaning only to the energy $E_\lambda$, keeping it finite
(and mostly small) throughout the computations.
Of course one can always consider $\lambda\sim\ell_p$ as a (limiting)
case of particular interest.
Indeed, we will see in the next Section that one can obtain significant
corrections induced by such modes in a way which is phenomenologically
acceptable within the semi-classical treatment.
\par
One observes that the factorization of the wave-function
$\tilde\psi_\lambda$ (gravitational state) into $f_\lambda$
(fluctuations) and a specific $\psi_c$ (classical part) is not forced
by Eq.~(\ref{scc}) or any other equation following from the WDW
equation (\ref{wdw}).
In Refs.~\cite{brout,bv,bfv} it was rather determined by the implicit
assumption that $\expec{\hat H_M}$ is the dominant contribution in
the semi-classical limit.
This physical assumption takes mathematical form in the condition
(\ref{UV}) which leads to the definition of the classical momentum
$P_c$ in Eq.~(\ref{pi}).
However, since in the last Section we concluded that there are times
at which $E_\lambda\sim\expec{\hat H_M}$ for every $\lambda$, this is
clearly contradictory and one should instead treat $E_\lambda$ as a
source for the dynamics of $R$ on the same footing as
$\expec{\hat H_M}$.
This can be achieved straightforwardly by introducing the modified
momentum
\be
P_\lambda=
-{1\over\kappa}\,{\partial R_\lambda\over\partial\eta_\lambda}
=-{1\over\kappa}\,\sqrt{2\,\kappa\,R_\lambda\,\expec{\hat H_M}
-\epsilon\,R_\lambda^2-\ell_p^4/\lambda^2}
\ .
\label{pil}
\ee
\par
A more formal way to derive this result is by defining a new WKB
wave-function $\psi_\lambda$ peaked on a modified trajectory
$R_\lambda$, parameterized by a time variable $\eta_\lambda$
($d\eta_\lambda=\eta_c\,d\eta_c$), such that
\be
\tilde\psi_\lambda=
{1\over\sqrt{-P_\lambda}}\,
e^{+{i\over\hbar}\,\strut\displaystyle\int P_\lambda\,dR_\lambda}
\,e^{i\,{R\over\lambda}}
\equiv\psi_\lambda\,\bar f_\lambda
\ .
\label{bwkb}
\ee
Then, upon substituting into Eq.~(\ref{wdw_g}),
in the classical adiabatic approximation $\dot R\ll R$
and neglecting the RHS, one obtains
\be
-i\,\kappa\hbar\,\psi_\lambda\,P\,
{\partial \bar f_\lambda\over\partial R}=
\psi_\lambda\,\left[-{\kappa\over 2}\,P^2-{\epsilon\over2\kappa}\,R^2
+R\,\expec{\hat H_M}
-{\kappa\hbar^2\over 2\,\lambda^2}\right]\,\bar f_\lambda
\ .
\label{sccc}
\ee
The LHS vanishes identically, since
$-\kappa\,\psi_\lambda\,P\,(\partial\bar f_\lambda/\partial R)
\equiv \partial\bar f_\lambda/\partial\eta_\lambda=0$.
Therefore the RHS gives the modified HJ equation
\be
-{1\over2\kappa}\,\left({d R_\lambda\over d\eta_\lambda}\right)^2
-{\epsilon\over2\kappa}\,R_\lambda^2
+R_\lambda\,\expec{\hat H_M}
-{\hbar\,\ell_p^2\over2\,\lambda^2}
=0
\ ,
\label{ehj}
\ee
which is Eq.~(\ref{pil}).
\subsection{Examples}
The above Eq.~(\ref{ehj}) will now be solved in the quantum adiabatic
approximation $\expec{\hat H_M}=N_\phi\,m_\phi$ constant and
for the three values taken by the parameter $\epsilon$ for the purpose
of showing explicit results.
However, we emphasize that the latter approximation is not essential
for the general formalism and is assumed just because it allows
to carry on the computation analytically.
\subsubsection*{Negative curvature}
For $\epsilon=-1$ the velocity $\dot R_\lambda$ vanishes at
\be
R_\lambda^\pm=-N_\phi\,{\ell_p^2\over\ell_\phi}\,\left[1
\pm\sqrt{1+{1\over N_\phi^2}\,{\ell_\phi^2\over\lambda^2}}
\right]
\ ,
\label{pm-1}
\ee
thus one expects a turning point at $R_\lambda=R_\lambda^-$
($R_\lambda^+<0$ is unphysical).
The latter reduces to the turning point $R_c^-=0$ for
$\ell_\phi/|\lambda|\to 0$.
Upon setting $R_\lambda(0)=R_\lambda^-$, the modified trajectory
is given by ($\eta\equiv\eta_\lambda$)
\be
R_\lambda=N_\phi\,{\ell_p^2\over\ell_\phi}\,\left[
\sqrt{1+{1\over N_\phi^2}\,{\ell_\phi^2\over\lambda^2}}\,\cosh\eta
-1\right]
\ ,
\label{rf-}
\ee
and the solution $R_c$ given in Eq.~(\ref{kc}) is recovered
as $R_\infty$ in the limit $\ell_\phi/|\lambda|\to 0$
($R_\lambda^-\to 0$).
In the opposite limit, $|\lambda|/\ell_p\to 0$, $R_\lambda^-$
diverges and the trajectory eventually reduces to a point.
\subsubsection*{Flat space}
For $\epsilon=0$ the modified trajectory is given by
\be
R_\lambda={\ell_p^2\over 2\,\lambda^2}\,{\ell_\phi\over N_\phi}
+{N_\phi\,\ell_p^2\over2\,\ell_\phi}\,\eta^2
\ ,
\label{rf0}
\ee
with a turning point at $R_\lambda(0)$.
The solution $R_c$ in Eq.~(\ref{kc}) is recovered in the limit
$\ell_p/|\lambda|\to 0$ ($R_\lambda(0)\to 0$) as for $\epsilon=-1$.
Also, the opposite limit behaves the same as for negative curvature.
\subsubsection*{Positive curvature}
For $\epsilon=+1$ there are two turning points at
\be
R_\lambda^\pm=N_\phi\,{\ell_p^2\over\ell_\phi}\,\left[1
\pm\sqrt{1-{1\over N_\phi^2}\,{\ell_\phi^2\over\lambda^2}}
\right]
\ ,
\ee
provided the square root is real, that is
$|\lambda|> {\ell_\phi/ N_\phi}$.
As before, for $\ell_\phi/|\lambda|\to 0$ the minimum
$R_\lambda^-\to R_c^-=0$ and the maximum
$R_\lambda^+\to R_c^+=2\,N_\phi\,\ell_p^2/\ell_\phi$.
Again, upon setting $R_\lambda(0)=R_\lambda^-$, the modified
solution is
\be
R_\lambda=N_\phi\,{\ell_p^2\over\ell_\phi}\,\left[1-
\sqrt{1-{1\over N_\phi^2}\,{\ell_\phi^2\over\lambda^2}}\,\cos\eta
\right]
\ .
\label{rf+}
\ee
Thus, the effect of the extra term in the HJ equation is to make
$R_\lambda$ oscillate between a minimum value $R_\lambda^-$ which is
shifted above zero and a maximum value which is below the turning
point $R_c^+$.
The shifts vanish and the solution $R_c$ given in Eq.~(\ref{kc})
is recovered in the limit $\ell_\phi/|\lambda|\to 0$.
At the opposite limit stands the case $|\lambda|=\ell_\phi/N_\phi$
for which the amplitude of the oscillation $R_\lambda^+-R_\lambda^-=0$.
\subsection{Consistency of the approximations}
\label{conditions}
The trajectories $R_\lambda$ displayed above are consistent only when
three different approximations hold simultaneously:
\begin{enumerate}
\item
\label{1}
The RHS's of Eqs.~(\ref{wdw_g}) and (\ref{wdw_m}) must be negligible.
This is the only relevant approximation which must hold for the general
formalism as developed in this Section to apply.
Since the RHS of Eq.~(\ref{wdw_g}) amounts to the expectation
value of a gravitational operator on the matter state $\tilde\chi$,
it is not substantially modified by the new definition of $\tilde\psi$
in Eq.~(\ref{ewkb}) with respect to the state (\ref{wkb})
(one only expects corrections because $\tilde\chi$ will evolve
differently with the new scale factor $R_\lambda$).
Hence, we refer the reader to Refs.~\cite{cv,bfv2,shell,cfv,fvv},
where this approximation has been studied at best for the case in
Eq.~(\ref{wkb}).
\par
However, in order to ensure that the corrections we have computed
are significant, we now need to check that the RHS of Eq.~(\ref{wdw_g})
is negligible with respect to the new term $|E_\lambda|$.
Taking the estimate for the RHS as given in Appendix B of
Ref.~\cite{cv},
\be
{\rm RHS}\sim{\hbar\,\ell_p^2\over R^3}\,N_\phi^2
\ ,
\ee
we obtain the condition
\be
{\rm RHS}\ll |E_\lambda|\ \ \Rightarrow\ \
R\gg N_\phi\,\lambda\sim N_\phi\,\ell_p
\ ,
\label{c_rhs}
\ee
where we keep on singling out the particular value
$\lambda\sim\ell_p$.
We can now take the trajectories given in the previous Section and
find that, for $\epsilon=0,-1$, the turning point
$R_{\lambda\sim\ell_p}(0)$ satisfies Eq.~(\ref{c_rhs}) provided
\be
N_\phi^2\ll{\ell_p^2\,\ell_\phi\over\lambda^3}
\sim{\ell_\phi\over\ell_p}
\ ,
\ee
while for $\epsilon=+1$ such a condition would be unphysical and is
never met.
This means that, for $\epsilon=+1$, the semi-classical approximation
breaks down before any rebound occurs and quantum fluctuations prevail
at small $R$ so that the superposition among different semi-classical
trajectories cannot be avoided.
Strictly speaking, the latter conclusion extends to all values of
$\epsilon$, since the WKB approximation breaks down near the classical
turning points
This actually turns out to be a blessing in disguise, since, in the
case of the collapse, the rebound of the sphere at a finite radius
inside the Schwarzschild radius would eventually violate causality
\cite{frolov}.
\par
A more complete analysis of the dynamics when the RHS's are not
negligible is currently under study \cite{forth}.
It is expected that one can still determine the evolution of the
system in this non-classical range (at least numerically), although
the geometrical interpretation is then lost (see, {\em e.g.},
Ref.~\cite{hajicek}).
\item
\label{2}
The quantum adiabatic approximation $\expec{\hat H_M}$ constant.
As we mentioned previously, this approximation is not essential for the
development of the general formalism and was actually relaxed in
Refs.~\cite{shell,cfv,fvv}.
However, it is only for $\expec{\hat H_M}$ constant that solutions can
be computed analytically.
Again we refer to Refs.~\cite{cv,bfv2,shell,cfv,fvv} for a detailed
analysis.
For the case of the gravitational collapse, one obtains the
condition \cite{cv}
\be
r-r(0)\gg\ell_\phi
\ ,
\label{Delta}
\ee
where $r\equiv \rho_s\,R_\lambda(\eta)$ is the areal radius of the
sphere and $r(0)$ is the turning point $\rho_s\,R_\lambda(0)$.
\item
\label{3}
The classical adiabatic approximation $\dot R_\lambda\ll R_\lambda$.
Despite the terminology, this approximation is not related to the
previous one, although, to some extent, it is inessential for the
general formalism developed here.
However, it is necessary for Eq.~(\ref{sc1}) for $f$ to simplify to the
tractable form (\ref{sc0}) and leads to a new condition which turns out
to be not really restrictive for the example worked out here.
In fact, it is easy to see that for the proposed solutions the
ratio
\be
{\dot R_\lambda\over R_\lambda}\to 0
\ ,
\ee
for $\lambda\to 0$ ($\epsilon=-1,0$) and $\lambda\to \ell_\phi/N_\phi$
(the minimum allowed value for $\epsilon=+1$).
\end{enumerate}
From the above considerations one thus concludes that the break-down
of the semi-classical approximation might occur at relatively large
values $\sim R_\lambda(0)$.
In the following Section, such turning/breaking points in the
gravitational collapse will be regarded as significantly different
from the point-like singularity only if they are bigger than the
Compton wave-length $\ell_\phi$ of the particles.
\setcounter{equation}{0}
\section{Applications}
\label{applications}
As described in Section~\ref{rev}, the contribution of gravitational
fluctuations incorporated in the theory is of quantum origin.
It is generally taken for granted that gravity in the world
we can test is classical, which leads one to assume the energy stored
in quantum gravitational fluctuations is negligible at the time
$\eta_0$ when measurements take place.
To be more precise, let us introduce the ratio
$\alpha^2(\eta)\equiv {|E_\lambda|/\expec{\hat H_M}}$
between quantum gravitational fluctuation energy and matter energy
and assume $\alpha_0\equiv\alpha(\eta_0)\ll 1$.
From Eq.~(\ref{Ela}) this definition can be used to express $\lambda$
as
\be
\lambda^2={1\over2\,\alpha^2_0}\,{\ell_p^2\,\ell_\phi\over R_0\,N_\phi}
\ ,
\label{la}
\ee
where $R_0=R_\lambda(\eta_0)$.
\par
The second important issue is whether the inclusion of $E_\lambda$
in the HJ equation leads to observable effects,
that is, one will have to check when (if ever) deviations from the
standard trajectories $R_c$ are physically significant in magnitude.
\par
In order to clarify the above points, we now specialize the very
simple analytic solutions found so far to two models.
In so doing we do not expect our results to be definite answers to
any basic physical question in either cases, however, we believe
they give hints as to the possible relevance of the predicted effects.
\subsection{Cosmology}
In the cosmological case one takes $\eta_0$ equal to the (conformal)
age of the Universe, so that the energy stored in the gravitational
fluctuations is totally negligible today.
However, this does not prevent $\alpha^2\sim 1/R_\lambda$ to be
comparable with one or bigger at very early stages.
The key observation is precisely that the present scale factor of
the universe, $R_0$, is related to the initial (minimum) scale factor
$R(0)$ by
\be
\alpha_0^2=\alpha^2(0)\,{R(0)\over R_0}
\ .
\label{00}
\ee
One also recalls that in the RW metric the spatial distance between
two points arbitrarily set at $\rho=0$ and at $\rho=\rho_d$ is given
by ($\rho_d\le 1$ for $\epsilon=0$)
\be
s=R\,\int_0^{\rho_d}{d\rho\over\sqrt{1-\epsilon\,\rho^2}}
\ .
\label{s-}
\ee
\subsubsection*{Open Universe}
When the spatial curvature $\epsilon=-1$, from Eqs.~(\ref{pm-1}) and
(\ref{la}) one obtains
\be
R(0)=R_\lambda^-=N_\phi\,{\ell_p^2\over\ell_\phi}\,\left[
\sqrt{1+2\,\alpha_0^2\,{R_0\,\ell_\phi\over\ell_p^2\,N_\phi}}
-1\right]
\ .
\label{r0-}
\ee
On using Eq.~(\ref{00}) one finds
\be
R(0)=2\,N_\phi\,{\ell_p^2\over\ell_\phi}\,\left(\alpha^2(0)-1\right)
\ ,
\ee
so that $R(0)>0$ implies $\alpha^2(0)>1$ and quantum gravitational
fluctuations must dominate the early stages in order to have a start
at non-zero scale factor (this is due to the gravitational potential
contributing with the same sign as $\expec{\hat H_M}$ in the HJ
equation for $\epsilon=-1$).
\par
It is now interesting to relate $\alpha_0$ to physically meaningful
quantities.
For instance, the relative difference $\Delta s=s_f-s_c$
between spatial distances measured when $R=R_\lambda(\eta_0)$ and
$R=R_c(\eta_0)$ is
\be
{\Delta s_0\over s_0}\simeq
\sqrt{1+2\,\alpha_0^2\,{R_0\,\ell_\phi\over\ell_p^2\,N_\phi}}
-1
\ ,
\ee
where we used Eqs.~(\ref{kc}) and (\ref{rf-}) in the limit
$\cosh\eta_0\gg 1$
($\Delta s_0>0$ since $R_\lambda\sim R_c+R_\lambda^-$).
Hence Eq.~(\ref{r0-}) becomes
\be
R(0)\simeq N_\phi\,{\ell_p^2\over\ell_\phi}\,{\Delta s_0\over s_0}
\ .
\ee
If one takes for $\Delta s_0/s_0$ the accuracy with which distances
are measured in the present Universe, the above equation gives us a
(very rough indeed) estimate of the maximum value for the initial
scale factor which cannot be ruled out by present measurements.
\par
A further consequence of having $R(0)>0$ is that two points in space
were causally disconnected at $\eta=0$ if their distance
$s(0)\gg\ell_\phi$, that is, from Eq.~(\ref{s-}),
\be
\rho_d\gg\rho_c=
\sinh\left({\ell_\phi\over N_\phi\,\ell_p^2}\,{s_0\over \Delta s_0}
\right)
\ .
\ee
\par
In the flat case $\epsilon=0$, $\alpha^2(0)=1$, as follows from a
quick inspection of Eq.~(\ref{ehj}), and the expression of the
initial scale factor simplifies to
\be
R(0)=\alpha_0^2\,R_0={\Delta s_0\over s_0}\,R_0
\ .
\ee
Then the causal comoving radius
$\rho_c={\ell_\phi\over R_0}\,{s_0\over \Delta s_0}$.
\subsubsection*{Closed Universe}
For $\epsilon=+1$ we have shown in Section~\ref{conditions} that our
corrections are less relevant, however, we also consider this case
for completeness.
On taking $R_0\sim R_\lambda^+$, one has
\be
R_0\simeq
2\,N_\phi\,{\ell_p^2\over\ell\phi}\,\left(1-\alpha_0^2\right)
\ ,
\ee
which holds for $\alpha_0^2<1$ (the gravitational potential in the
HJ equation is now opposite in sign to $\expec{\hat H_M}$).
The relative difference in distances is
\be
{\Delta s_0\over s_0}\simeq -\alpha_0^2
\ ,
\ee
where we used Eqs.~(\ref{kc}) and (\ref{rf+}) in the limit
$\cos\eta_0\simeq-1$ ($\Delta s_0<0$ because the maximum of
$R_\lambda$ is shifted down with respect to the maximum of $R_c$).
Putting the pieces together gives
\be
R(0)=\left|{\Delta s_0\over s_0}\right|\,R_0
\ ,
\ee
and $s(0)\gg\ell_\phi$ if
\be
\rho\gg\rho_c=\sin\left(
{\ell_\phi\over R_0}\,\left|{s_0\over \Delta s_0}\right|\right)
\ .
\ee
Since $\rho<1$, the latter condition can be satisfied only if
\be
\left|{\Delta s_0\over s_0}\right|\gg{2\,\ell_\phi\over\pi\,R_0}
\ .
\ee
\par
No need to say the present model is too simplified to take the
expression for the comoving radius $\rho_c$ seriously as a
prediction for the scale of Cosmic Background Fluctuations or
related cosmological quantities.
In fact, there is no inflationary stage and regions outside of
$\rho_c$ will eventually come into the causal cone after a finite
(short) time due to the slow expansion of the scale factor in the
FRW models.
The situation might change in case one considers a more realistic
description.
Further, we recall that the existence of a minimum scale factor
$R(0)>0$ is a basic ingredient of {\em Pre-Big-Bang Cosmology}
(see, {\em e.g.} \cite{ve} and Refs. therein), in which case
$\phi$ should be identified with the homogeneous mode of the
dilaton predicted by the low energy limit of string theory
\cite{bob}.
\subsection{Gravitational collapse}
For the case of the collapsing sphere of dust the above framework is
inverted since now $|E_\lambda|$ increases along the classical
trajectory.
Thus, although one starts with $\alpha_0\ll 1$ so that the energy of
the quantum gravitational fluctuations is totally negligible,
when the singularity is approached the gravitational fluctuations
induce an effective quantum pressure which slows down the collapse
and causes the break-down of the semi-classical approximation at a
finite radius of the sphere $r(0)$.
A very important observation, already mentioned at the end of
Section~\ref{EHJ}, is that $r(0)$ is physically distinguished
from the singularity $\rho_s\,R_c=0$ only if it is bigger than the
Compton wavelength $\ell_\phi$.
\par
Before proceeding, it is useful to recall that the Schwarzschild
radius of the sphere $r_H=2\,M$ where the ADM mass parameter is
\be
M=\rho_s^3\,N_\phi\,{\ell_p^2\over\ell_\phi}
\ ,
\label{M}
\ee
regardless of the value of $\epsilon$.
We will assume $M/\kappa$ is the mass that is measured for
astronomical objects, although it generally differs from the
proper mass
$N_\phi\,(\hbar/\ell_\phi)\,\int_0^{\rho_s}
{\rho^2\,d\rho/\sqrt{1-\epsilon\,\rho^2}}$.
Further, $\rho_s$ is related to the geodesic energy parameter
${\cal E}$ of the trajectory $r_s=\rho_s\,R$ of the radius of the
sphere in the outer Schwarzschild space-time with mass parameter $M$
by ${\cal E}^2=1-\epsilon\,\rho_s^2$
($-1<{\cal E}<1$ for bound orbits, ${\cal E}\ge 1$ for unbound orbits)
\cite{stephani}.
\subsubsection*{Scattering orbits}
For ${\cal E}>1$ one can choose the starting radius of the sphere
$r_0=\rho_s\,R_0$ is any value greater than $r_H$.
Then the sphere will bounce in correspondence to $R_\lambda^-$ at
\be
r(0)={M\over {\cal E}^2-1}\,\left[
\sqrt{1+2\,({\cal E}^2-1)\,\alpha_0^2\,{r_0\over M}}-1
\right]
\ .
\label{r-}
\ee
As mentioned above, $r(0)$ must be greater than
$\ell_\phi$ to be physical, that is
\be
\alpha_0^2\gg {\ell_\phi\,[r_H+({\cal E}^2-1)\,\ell_\phi]\over
r_0\,r_H}
\ .
\ee
For ${\cal E}^2-1$ small one can expand the square root in
Eq.~(\ref{r-}) and obtains
\be
r(0)\simeq \alpha_0^2\,r_0
\ ,
\label{10}
\ee
with the condition $\alpha_0^2\gg {\ell_\phi/r_0}$.
The above result is exact for ${\cal E}=1$, in which
case $\rho_s$ is arbitrary and $r=(\alpha_0^2\,r_0)+\rho_s\,R_c$.
In the opposite limit, ${\cal E}\gg 1$, one obtains
\be
r(0)\simeq {\alpha_0\over{\cal E}}\,\sqrt{r_0\,r_H}
\ ,
\ee
with $\alpha_0^2\gg {\cal E}^2\,\ell_\phi^2/r_H\,r_0$.
\par
In all cases ($\epsilon=-1,0$), $r(0)$ is bigger than $r_H$ provided
$\alpha_0^2>r_H\,{\cal E}^2/r_0$, or, using Eq.~(\ref{la}) and
(\ref{M}), the comoving radius is given by
\be
{\rho_s^2\over\sqrt{1-\epsilon\,\rho_s^2}}\sim {2\,M\over\ell_p}
\ ,
\ee
for the limiting case $\lambda\sim\ell_p$.
\subsubsection*{Bound orbits}
For ${\cal E}^2<1$, on setting $r_0=\rho_s\,R_\lambda^+$,
the choice (\ref{la}) and the trajectory (\ref{rf+}) give
\be
r_0={r_H\over 1-{\cal E}^2}\,\left(1-\alpha_0^2\right)
\ .
\ee
Then, the radius at which the sphere bounces is given by
\be
r(0)={M\over 1-{\cal E}^2}\,\left[1
-\sqrt{1-{4\,\alpha_0^2}\,\left(1-{\alpha_0^2}\right)}\right]
\simeq{\alpha_0^2}\,r_0
\ ,
\label{bounce}
\ee
and one concludes that Eq.~(\ref{10}) holds for $-1<{\cal E}\le 1$,
in which cases $r(0)\gg\ell_\phi$ provided
$\ell_\phi/r_0\ll \alpha_0^2\ll 1$
and $r(0)>r_H$ for $r_h/r_0< \alpha_0^2\ll 1$.
For the limiting value $\lambda\sim\ell_p$ this means
\be
\rho_s^2\sim {2\,M\over\ell_p}\gg 1
\ ,
\ee
which lies outside the allowed range of $\rho_s$ and the breaking
point cannot thus be bigger than $r_H$ for $\epsilon=+1$.
\par
It is worth noting that the minima obtained above are fairly generic
in that they do not depend on the detailed structure of the sphere
nor on the specific form of the quantum gravitational fluctuations
($\ell_\phi$, $\ell_p$ and $\lambda$ do not appear explicitly).
Furthermore, the turning point at which the
semi-classical approximation breaks can be rather big and, from
Eq.~(\ref{10}), one cannot exclude it occurs
at a radius comparable with the error with which one measures $r_0$.
Of course, to model a star with a sphere of dust ignores
(among the rest) the crucial role played by the pressure in keeping
the star in equilibrium and contrasting the collapse itself.
Hence, it is clear that the actual value of $r(0)$ could be
significantly different from the one estimated here.
\par
We conclude by mentioning an independent argument which supports
the possibility of having the kind of pressure emerging from the
quantum fluctuations discussed in this Section.
In fact, besides the WDW equation, one has the conservation of the
total energy of the system, namely its ADM mass \cite{madm}
\be
M=\rho_s^3\,\left(\expec{\hat H_M}-E_f\right)
\simeq\rho_s^3\,\expec{\hat H_M(\eta_0)}
\ .
\label{Madm}
\ee
Since $\expec{\hat H_M}$ increases in time, due to non-adiabatic
production of matter particles \cite{shell,cfv} (an effect totally
ignored in the present notes), Eq.~(\ref{Madm}) requires
that either $\rho_s$ decreases or $E_f$ increases
(or both effects take place).
The first case amounts to {\em quantum jumps} to classical
trajectories with geodesic energy closer and closer to ${\cal E}=1$
which would act as a {\em semi-classical attractor} \cite{rc}.
The second case would imply that, although one can start with a state
in which only matter modes are present (as appears sensible for a
sphere of large initial radius), the price to pay for preserving
the classical dynamics is the generation of gravitational
perturbations in an amount such that the total energy
(along with ${\cal E}$) is conserved.
We remark that the spherical symmetry assumed for the model
would prevent these perturbations from propagating in the external
vacuum as gravitational waves ({\em Birkhoff's theorem}).
\setcounter{equation}{0}
\section{Conclusions}
\label{conc}
We have generalized the BO approach to the WDW equation in FRW
minisuperspace \cite{bfv} in order to include homogeneous quantum
gravitational fluctuations around the WKB trajectory.
In a standard approach, such fluctuations are treated as
perturbations of a fixed background and satisfy a Schr\"odinger
equation, whose solutions in (double) adiabatic approximation were
displayed both in Lorentzian and Euclidean space.
One then realizes that these solutions signal a possible instability,
since their energy can grow without bound for small $R$.
The latter result suggested that the semi-classical limit
had been incorrectly identified.
Then a second approach was proposed in which the semi-classical
limit includes the back-reaction of quantum gravitational
fluctuations on the metric from the start, in much the same fashion
as in quantum field theory on curved back-ground one replaces the
classical energy-momentum tensor of matter with the expectation
value of the corresponding quantum operator (for matter and
gravitational waves).
This in fact led to different classical trajectories with a
 possible
non-vanishing minimum size for a FRW Universe or a collapsing body
at which the semi-classical approximation breaks down and the metric
does not admit a description in terms of classical variables.
\par
Although the FRW minisuperspace model is too simple to be realistic,
one learns that the role played by quantum gravitational fluctuations,
at least, should not be overlooked when considering self-gravitating
matter.
The guiding analogy is the treatment of infrared divergences in
quantum field theory.
In that context one encounters diverging quantities when studying,
{\em e.g.,} the Bremsstrahlung from an electron moving in the external
electric field of a nucleus.
At the classical level (the tree level of the quantum theory)
the transition amplitude diverges for vanishing energy of the emitted
photon.
This problem is cured by adding the (diverging) one-loop contribution
and noting that experimental measurements would not distinguish
the final state of the electron with energy $E$ from any state with
energy $E-\Delta E$ if $\Delta E$ is smaller than the precision
$\Lambda$ of the apparatus.
Then one has to sum over all the (tree level) emissions of (soft)
photons of energy $\Delta E<\Lambda$ and obtains the counter-term
which precisely cancels against the one-loop diverging term.
Perhaps one can rephrase the results obtained in these notes by saying
that the inclusion of (soft) quantum gravitational fluctuations with
energy smaller than the precision with which we measure the energy
of matter seems to cure the singularity in the density distribution
of matter which develops at the classical level according to General
Relativity.
\subsubsection*{Acknowledgments}
I would like to thank G. Venturi for suggesting the problem and useful
discussions.

\begin{thebibliography}{99}
\bibitem{opposite}
Had we chosen the opposite sign in the definition of the time
(\ref{d_eta}) the matter Hamiltonian $\expec{\hat H_M}<0$ and
$E_\lambda>0$.
In any case the matter and gravitational contributions enter with
opposite signs in the total Hamiltonian.
\bibitem{madm}
Eq.~(\ref{Madm}) follows straightforwardly from the
matching conditions between the internal Robertson-Walker metric and
the external Schwarzschild metric \cite{cv} or, equivalently, from
the Einstein-Hilbert action of the whole system once the symmetry
and the matter content are specified \cite{rc}.
\end{thebibliography}
\end{document}